\documentclass[journal]{IEEEtran}
\usepackage{cite}

\ifCLASSINFOpdf
   \usepackage[pdftex]{graphicx}
   \usepackage{color}
   \usepackage{epstopdf}
\else
\fi
\usepackage{amsmath}
\usepackage{bbm}
\usepackage{bm}                       
\usepackage{amssymb}                  

\usepackage{algorithm}
\usepackage{algorithmic}
\begin{document}
%
\title{\huge Optimized Trajectory Design in UAV Based Cellular Networks for 3D Users: A Double Q-Learning Approach}
%
%
%

\author{Xuanlin~Liu,~Mingzhe~Chen,~and~Changchuan~Yin%
\thanks{X. Liu, M. Chen, and C. Yin were with the Beijing Laboratory of Advanced Information Network, and the Beijing Key Laboratory of Network System Architecture and Convergence, Beijing University of Posts and Telecommunications,
Beijing 100876, China (e-mail: xuanlin.liu@bupt.edu.cn; chenmingzhe@bupt.edu.cn; ccyin@ieee.org).}
\thanks{This paper has been accepted by Journal of Communications and Information Networks, February 1, 2019.}}
\maketitle

\begin{abstract}
In this paper, the problem of trajectory design of unmanned aerial vehicles (UAVs) for maximizing the number of satisfied users is studied in a UAV based cellular network where the UAV works as a flying base station that serves users, and the user indicates its satisfaction in terms of completion of its data request within an allowable maximum waiting time. The trajectory design is formulated as an optimization problem whose goal is to maximize the number of satisfied users. To solve this problem, a machine learning framework based on double Q-learning algorithm is proposed. The algorithm enables the UAV to find the optimal trajectory that maximizes the number of satisfied users. Compared to the traditional learning algorithms, such as Q-learning that selects and evaluates the action using the same Q-table, the proposed algorithm can decouple the selection from the evaluation, therefore avoid overestimation which leads to sub-optimal policies. Simulation results show that the proposed algorithm can achieve up to 19.4\% and 14.1\% gains in terms of the number of satisfied users compared to random algorithm and Q-learning algorithm.
\end{abstract}


%
\IEEEpeerreviewmaketitle

\section{Introduction}
Due to its flexible mobility and low cost, unmanned aerial vehicle (UAV) communication is viewed as an important solution in future communication systems \cite{Zeng2016Wireless}. In fact, UAVs have already been considered to be deployed in many fields \cite{Xu2018UAV,8614433}, such as wireless power transfer, secure communications, relaying, wireless sensor networks, and caching. However, applying UAVs in communication systems still faces many challenges, such as deployment, effective resource allocation, energy efficiency, and trajectory design.

The existing literature has studied a number of problems related to the use of UAVs for wireless communication such as \cite{8247211,7412759,Chen2017Caching,Hu2018UAV,zhou2018computation,yang2018joint,Yuan2018,8432474}. The work in \cite{8247211} considered a multi-UAV enabled wireless communication system, where multiple UAV-mounted aerial base stations are employed to serve a group of users. The authors in \cite{7412759} investigated the deployment of UAVs as flying base stations so as to provide flying wireless communications to certain geographical area.
In \cite{Chen2017Caching}, the authors proposed a UAV based framework to provide service for the mobile users in a cloud radio access network system. The work in \cite{Hu2018UAV} deployed UAVs in a cellular network and designed optimal spectrum trading, so as to provide temporarily downlink data offloading. In \cite{zhou2018computation}, the authors proposed a UAV-enabled mobile edge computing system for maximization of computation rate. The authors in \cite{yang2018joint} investigated an uplink power control problem for UAV based wireless networks. The work in \cite{Yuan2018} analyzed the link capacity between autonomous UAVs with random trajectories. In \cite{8432474}, the authors studied the capacity region of a UAV-enabled two-user broadcast channel.
However, most of the existing literature such as \cite{Hu2018UAV,7412759,Chen2017Caching,8247211,zhou2018computation,yang2018joint,Yuan2018,8432474} that uses the UAVs as high-altitude, static base stations or relays doesn't consider the flexible mobility of UAVs which can provide service for the ground users. Moreover, these existing works only focus on the ground users and do not consider to provide service to the users that are located in the air. Indeed, none of this existing body of literature analyzes the potential of using machine learning tools for leveraging the movable nature of UAV to assist wireless communications. The complexity of UAV trajectory design makes it essential to introduce reinforcement learning algorithm to optimize the performance of UAV-assisted wireless networks.


The use of reinforcement learning for solving communication problems was studied in \cite{Xiao2018RL,Toumi2018D2D,Hu2017LTEU,8395443,7588197,8410619,Challita2018AI}. The work in \cite{Xiao2018RL} applied deep Q-network in a mobile communication system to reduce the exploration time for achieving an optimal communication policy. The authors in \cite{Toumi2018D2D} proposed a Q-learning based algorithm to coordinate power allocation and control interference levels, so as to maximize the sum data rate of device-to-device (D2D) users while guaranteeing QoS for cellular users. In \cite{Hu2017LTEU}, the authors proposed an expected Q-learning algorithm to solve the spectrum allocation problem in LTE networks that operate in unlicensed spectrum (LTE-U) with downlink-uplink decoupling and improve the total rate. In \cite{8395443,7588197}, the authors proposed an echo state network (ESN) based learning algorithm to solve the spectrum allocation problem in wireless networks. The work in \cite{8410619} proposed a reinforcement learning scheme to improve spectral efficiency in cloud radio access networks. The authors in \cite{Challita2018AI} used the artificial neural network to provide a reliable wireless connectivity for cellular-connected UAVs.
However, most of the existing works \cite{Xiao2018RL,Toumi2018D2D,Hu2017LTEU,8395443,7588197,8410619,Challita2018AI} focused on the use of traditional Q-learning algorithm but ignored its disadvantage, i.e., the traditional Q-learning algorithm has large overestimations of the values. In UAV based wireless networks, the overestimation will result in a sub-optimal design of the UAV trajectory or a sub-optimal policy of resource allocation, which will degrade the performance of wireless networks.

The main contribution of this paper is to develop a novel framework that enables UAVs to find an optimal flying trajectory to maximize the number of satisfied users in a cellular network. \textit{To our best knowledge, this is the first work that considers the flying trajectory of UAVs with the three dimensional users that have their own data requests and delay requests.} In this regard, our key contributions are summarized as follows:
\begin{itemize}
  \item We propose a novel model of UAV based cellular network where the UAV is deployed as a flying and movable base station for downlink transmission. In this model, users are divided into ground users and aerial users. All of the users will send their data request and delay request to the UAV and the UAV will design an optimal flying trajectory to maximize the number of satisfied users. 
  \item We develop a double Q-learning framework to optimize the flying trajectory of the UAV so as to maximize the number of satisfied users. Compared to the traditional Q-learning algorithm \cite{Bennis2010Q}, the proposed algorithm uses two Q-tables to decouple the selection from the evaluation in case of overestimation caused by selecting and evaluating actions in only one Q-table. Hence, the proposed algorithm can converge to the optimal trajectory which leads to the maximum of satisfied users.
  \item Simulation results show that, in terms of the number of satisfied users, the proposed algorithm can yield up to 19.4\% gain compared to the random algorithm and 14.1\% gain compared to Q-learning algorithm.
\end{itemize}

The rest of this paper is organized as follows. The system model and problem is described in Section~\ref{sec:model}. The double Q-learning based optimal trajectory design is proposed in Section~\ref{sec:al}. In Section~\ref{sec:sim}, numerical simulation results are presented and analyzed. Finally, conclusions are drawn in Section~\ref{sec:con}.

\section{System Model and Problem Formulation}\label{sec:model}
\begin{figure}[!t]
\centerline{\includegraphics[width=3in]{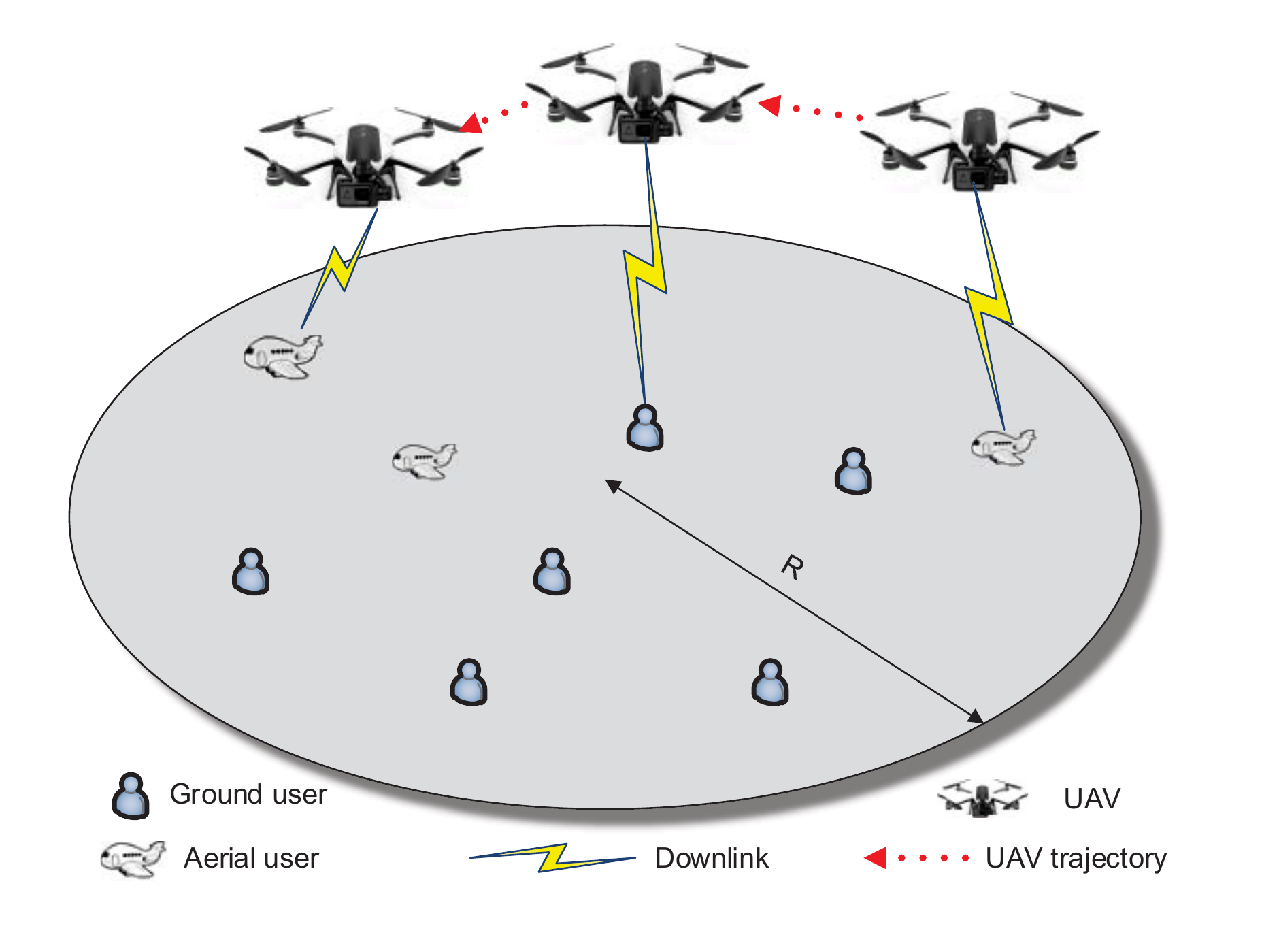}}
\caption{A cellular network with a UAV base station.}
\label{figure:model}
\end{figure}
Consider the downlink transmission of a cellular network that consists of an unmanned aerial vehicle (UAV) serving a set $\mathcal{U}=\left\{1,2,\ldots,U\right\}$ of $U$ users shown in \figurename~\ref{figure:model}. 
In our model, we consider two type of users: ground users and aerial users. The ground users are the traditional cellular users standing on the ground, which is denoted by a set $\mathcal{G}=\left\{1,2,\ldots,U_{\rm G}\right\}$ of {\color{black}$U_{\rm G}$} ground users. The aerial users represent the cellular users in the air such as camera drones, sensor drones, and aerial vehicles, which is denoted by a set $\mathcal{A}=\left\{1,2,\ldots,U_{\rm A}\right\}$ of $U_{\rm A}$ aerial users.
{\color{black}Service operators provide all information related to the users to the UAV, such as users' data requests and locations.} The location of each user $i$ is considered as a three dimensional coordinate, which is denoted by $\left(x_i,y_i,h_i\right)$, where $x_i$ and $y_i$ are the horizontal coordinates, and $h_i$ is the altitude of each user $i$. For ground user $i$, the altitude will be $h_i=0$. For aerial user $i$, the altitude $h_i$ is typically greater than 50 meters.
The UAV will fly from a designated position to serve each user at a fixed altitude $H$. In the studied model, the UAV can only serve one user at each time slot and the UAV will only fly to serve the next user after the previous service is completed.

Each user's request that is sent to the UAV consists of two elements: the size of data that each user requests and the maximum time that the user can wait for service, which is referred as endurance time. Mathematically, let $A_i$ be the size of the data that each user requests and $T_i$ be the endurance time. The size of the data that each user requests depends on the service type.
Meanwhile, the endurance time consists of the time that each user waits for the UAV to arrive and the time that the UAV serves the user. If the request of a given user is completely served before the endurance time, the user will be satisfied with the service provided by the UAV. The user that satisfies the service provided by the UAV is referred as a satisfied user. Next, we first introduce the waiting delay and the time that UAV serves the user. Then, we formulate the problem of maximizing the number of satisfied users.

\subsection{Waiting Delay}
The waiting delay of each user $i$ consists of the time that the UAV serves the users before user $i$ and the time that the UAV flies to user $i$. We assume that the speed of {\color{black} the UAV} is $c_{\rm U}$ and the distance between {\color{black} the UAV} and user $i$ is $\color{black}d_{i}$. The flying time from {\color{black} the UAV} to user $i$ is given by: 
\begin{equation}
\color{black}
t^\textrm{F}_i=\frac {d_{i}}{c_{\rm U}}.
\end{equation}

{\color{black}We assume that the time that the UAV starts to fly to user~$i$ is given by:}
 \begin{equation}
  \color{black}
  t^\textrm{S}_i\left(\boldsymbol{e} \right)=t_{i^{\rm Pre}}\left(\boldsymbol e\right),
 \end{equation}
{\color{black}where $t_{i^{\rm Pre}}\left(\boldsymbol e\right)$, which will be further defined in Subsection~\ref{subsec:transd}, represents the total service time for user $i^{\rm Pre}$, $i^{\rm Pre}$ represents the user that is served before user $i$, and $\boldsymbol{e}=\left[{e_1,\ldots, e_U}\right]$ represents users' service order with $e_i=k, k \in \mathcal{U}$}. $e_i=k$ represents that user $i$ is served at order $k$. For example, $e_i=1$ denotes that {\color{black} the UAV} will serve user $i$ first. Note that, if user $i$ is served first, $t_i^\textrm{S}=0$. $t^\textrm{S}_i$ depends on the number of users that have already been served by {\color{black} the UAV}. The total waiting time of each user $i$ is given by:
\begin{equation}\label{eq:waiting}
t^\textrm{W}_i \left(\boldsymbol{e} \right)=t^\textrm{F}_i+t^\textrm{S}_i \left(\boldsymbol{e} \right).
\end{equation}

From \eqref{eq:waiting}, we can see that the total waiting delay for user~$i$ depends on both the flying time and the total service time for the previous served user $i^{\rm Pre}$. We can also see that as the service order $\boldsymbol e$ changes, the time that {\color{black} the UAV} finishes the previous service and the total waiting delay also change.


\subsection{Transmission Delay}\label{subsec:transd}
Next, we introduce the models of transmission links between UAV and users. Due to the altitude differences between ground users and aerial users, the channel conditions are also different. In consequence, the UAV-ground user and UAV-aerial user transmission links are defined separately as follows.

{\it 1) UAV-Ground User Links:} The probabilistic UAV channel model is used to model the transmission link between {\color{black} the UAV} and ground user $i$. Probabilistic line-of-sight (LoS) and non-line-of-sight (NLoS) links are considered in \cite{7412759}, which explains that the attenuation in NLoS is much higher than that in LoS link due to the shadowing and diffraction loss. The LoS and NLoS channel gains of {\color{black} the UAV} transmitting data to ground user $i$ is given by \cite{Al2014Modeling}:
\begin{equation}
\color{black}
  \begin{aligned}
  g_{i}^\textrm{LoS}=d_{i}^{-\alpha},~~~~
  g_{i}^\textrm{NLoS}=\eta_{\rm NLoS} d_{i}^{-\alpha},
  \end{aligned}
\end{equation}
respectively, where $\alpha$ is the path loss exponent for the UAV transmission link, and $\eta_{\rm NLoS}$ is an additional attenuation factor caused by the NLoS connection. According to \cite{Al2014Modeling}, the probability of the LoS link is given by:
\begin{equation}
\color{black}
  \textrm{Pr}\left(g_{i}^\textrm{LoS}\right)=\left(1+X\exp\left(-Y\left[\phi_{i}-X\right]\right)\right)^{-1},
\end{equation}
where $X$ and $Y$ are environmental parameters and $\color{black}\phi_{i}=\sin^{-1}\left(\Delta_{i}/d_{i}\right)$ is the elevation angle with $\color{black}\Delta_{i}=\left|H-h_i \right|$ being the deviation between the user $i$'s altitude $h_i$ and the UAV's fixed altitude $H$. Then the average channel gain from {\color{black} the UAV} to ground user $i$ is given by \cite{Al2014Modeling}:
\begin{equation}
\color{black}
 {{g}_{i}^{\rm G}}=\textrm{Pr}\left(g_{i}^\textrm{LoS}\right)\times g_{i}^\textrm{LoS}+\textrm{Pr}\left(g_{i}^\textrm{NLoS}\right)\times g_{i}^\textrm{NLoS},
\end{equation}
where $\color{black}\textrm{Pr}\left(g_{i}^\textrm{NLoS}\right)=1-\textrm{Pr}\left(g_{i}^\textrm{LoS}\right)$. The downlink rate of ground user $i$ is given by:
\begin{equation}
\color{black}
  c_{i}^{\rm G}=B\log_2\left(1+\gamma_{i}^{\rm G}\right),
\end{equation}
where $\color{black}\gamma_{i}^{\rm G}=\frac{P{{g}_{i}^{\rm G}}}{\sigma ^2}$ is the downlink signal-to-noise ratio (SNR) between {\color{black} the UAV} and ground user $i$, $B$ is the bandwidth of the downlink transmission links, $P$ is the transmit power of the UAV, and $\sigma ^2 $ is the variance of the Gaussian noise. 

{\it 2) UAV-Aerial User Links:} {\color{black}The millimeter wave (mmWave) propagation channel is used to model the transmission link between {\color{black} the UAV} and aerial user $i$. The mmWave channel can provide high transmission rate so that the UAV can finish users' requests quickly and in time.} Due to the high altitudes of {\color{black} the UAV} and aerial user $i$, the transmission link can be considered as a LoS link, whose path loss is given by \cite{Chen2017Caching,Rap2013mmWave} in dB:
\begin{equation}
  \color{black}
  p_{i}^\textrm{A}\left({\rm dB}\right)=20\log\left(\frac{4\pi d_{i}f}{c}\right)+\eta_\textrm{LoS},\\
\end{equation}
where $\color{black}20\log\left(\frac{4\pi d_{i}f}{c}\right)$ represents the free space path loss with $\color{black}d_{i}$ being the distance between aerial user $i$ and {\color{black} the UAV}, $f$ is the carrier frequency of mmWave and $c$ represents the speed of light, $\eta_\textrm{LoS}$ represents the additional attenuation factor due to the LoS connections. The downlink rate of aerial user $i$ is given by:
\begin{equation}
\color{black}
  c_{i}^{\rm A}=B\log_2\left(1+\gamma_{i}^{\rm A}\right),
\end{equation}
where $\color{black}\gamma_{i}^{\rm A}=\frac{P}{{{p}_{i}^{\rm A}}\sigma ^2}$ is the downlink signal-to-noise ratio (SNR) between {\color{black} the UAV} and aerial user $i$.

\vspace{1em}
Hence, the transmission delay of serving user $i$ is given by:
\begin{equation}
\color{black}
  t_i^\textrm{T}=\left\{\begin{array}{cc}
    \frac{A_i}{c_{i}^{\rm G}},&\textrm{~~for ground users},\\
    \frac{A_i}{c_{i}^{\rm A}},&\textrm{for aerial users}.
    \end{array}\right. 
\end{equation}

The total time that UAV serves user $i$ can be calculated as:
\begin{equation}\label{eq:totald}
 t_i \left(\boldsymbol{e} \right)=t_i^\textrm{T}+t_i^\textrm{W} \left(\boldsymbol{e} \right). 
\end{equation}

From \eqref{eq:totald}, we can see that the total time that UAV serves user $i$ depends on both the waiting delay and transmission delay. The waiting delay changes as the service order $\boldsymbol e$ changes in \eqref{eq:waiting}. In consequence, the total time that UAV serves user $i$ also changes.


\subsection{Problem Formulation}
Given the defined system model, our goal is to design a flying trajectory so as to maximize the number of satisfied users. Next, we first introduce the notion of the satisfied user. Then, we formulate the optimization problem. 
Given the size of the data that user $i$ requests $A_i$ and the endurance time $T_i$, the satisfaction indicator of user $i$ is defined as follows:
\begin{equation}\label{eq:s}
\color{black}
  f_{i} \left(\boldsymbol{e} \right)=\mathbbm{1}_{\left\{ t_i \left(\boldsymbol{e} \right)  \leqslant  T_i  \right\}}=\mathbbm{1}_{\left\{ t^\textrm{F}_i+t^\textrm{S}_i \left(\boldsymbol{e} \right)+t_i^\textrm{T} \leqslant  T_i  \right\}},
\end{equation}
where $\mathbbm{1}_{\left\{ x  \right\}}=1$ as $x$ is true; otherwise, $\mathbbm{1}_{\left\{x \right\}}=0$. $\color{black}f_{i}\left(\boldsymbol e\right)=1$ indicates that, under the service order $\boldsymbol e$, {\color{black} the UAV} completes the request of user $i$ within the endurance time and user $i$ is satisfied.

Having introduced the notation of a satisfied user in \eqref{eq:s}, the next step is to introduce a flying trajectory management mechanism for {\color{black} the UAV} to maximize the number of satisfied users. This problem can be formulated as:
\begin{subequations}\label{eq:max}
\begin{equation}
\color{black}
\mathop {\max }\limits_{\boldsymbol{e}}  \sum\limits_{i \in \mathcal{U}}f_{i} \left(\boldsymbol{e} \right),\label{subeq:max_1}
\end{equation}
\vspace{-2em}
\begin{equation}
{\rm s.\;t.}~~ e_{i} \in \mathcal{U},
\end{equation}
\end{subequations}
where $\color{black}f_{i} \left(\boldsymbol{e} \right)$ is defined in \eqref{eq:s} and $e_i$ represents the service order of user $i$. The problem \eqref{eq:max} aims to find the optimal trajectory so that the UAV can complete most of the users' requests within their endurance time after receiving all the user requests. Since finding the optimal trajectory needs to evaluate all possible permutations of service order $\boldsymbol e$, which takes up a substantial amount of service time, it's essential to introduce a learning algorithm to shorten the calculation time for the trajectory.

\section{Double Q-Learning Framework For Maximizing the Number of Satisfied Users}\label{sec:al}
To solve the maximization problem in \eqref{eq:max}, we introduce a reinforcement learning framework based on double Q-learning. Compared to the existing reinforcement learning algorithms \cite{Xiao2018RL,Toumi2018D2D,Hu2017LTEU} such as Q-learning that may result in sub-optimal trajectory and leads to the number of satisfied users not maximized, the proposed double Q-learning algorithm enables the UAV to find the optimal flying trajectory to serve the users so as to maximize the number of satisfied users. Moreover, compared to the traditional Q-learning algorithm that typically uses one Q-table to record and update the values resulting from different states and actions \cite{Bennis2010Q}, the proposed double Q-learning algorithm uses two Q-tables to separately select and evaluate the actions. In this regard, the proposed double Q-learning algorithm avoids the overestimation of Q values. The overestimation usually occurs in traditional Q-learning algorithm due to the positive feedback caused by selecting and evaluating the action in the same Q-table.

Next, we first introduce the components of the double Q-learning algorithm. Then, we explain the procedure of the use of double Q-learning algorithm to find the optimal flying trajectory for the UAV.

\subsection{Components of Double Q-Learning Algorithm}
A double Q-learning model consists of four basic components: a) agent, b) actions, c) states, and d) reward function, which are specified as follows.
\begin{itemize}
  \item \textbf{Agent:} In this problem, the agent is obviously the UAV. The UAV can collect the users' information such as users' locations, the size of the data that users request and the endurance time. 
  \item \textbf{Action:} The actions of the double Q-learning algorithm determines the user that UAV will serve at next time slot. Let $\color{black}a$ be an action of {\color{black} the UAV} and $\color{black}a \in  \mathcal U$. For example, $\color{black}a =k$ denotes that {\color{black} the UAV} will provide service for user $k$. 
  \item \textbf{State:} Each state of {\color{black} the UAV}, $\color{black}\boldsymbol{s}=\left[\boldsymbol{w},\boldsymbol{v},\boldsymbol{t}^\textrm{S},\boldsymbol{t}^\textrm{F}\right]$ consists of: 1) the vector $\boldsymbol{w}=\left[w_1, \ldots, w_U\right]$ that denotes each user whether has been served by {\color{black} the UAV},  where $w_i \in \left\{0,1 \right\}$ with $w_i=1$ denotes that user $i$ has already been served by {\color{black} the UAV}, otherwise, $w_i=0$; 2) endurance time vector $\boldsymbol{v}=\left[{v_1,\ldots,v_U}\right]$ where $v_i=T_i$; 3) waiting time vector $\boldsymbol{t}^\textrm{S}=\left[ {t}_1^\textrm{S}, \ldots, {t}_U^\textrm{S}\right]$ with $t_{i}^S=0$ denotes that user $i$ has not been served by {\color{black} the UAV}; 4) flying time vector $\boldsymbol{t}^\textrm{F}=\left[ {t}_1^\textrm{F}, \ldots, {t}_U^\textrm{F} \right]$. 
  \item \textbf{Reward:} The reward function $\color{black}r\left(\boldsymbol{s},a\right)$ is defined as the total number of satisfied users as {\color{black} the UAV} takes action $\color{black}a$ under state $\color{black}\boldsymbol{s}$. $\color{black}r\left(\boldsymbol {s},a\right)$ can be specified as follows:
  \begin{equation}\label{eq:reward}
    \color{black}
    r\left(\boldsymbol{s},a_j\right)=\sum\limits_{i = 1}^U w_i .
  \end{equation}
\end{itemize}

\subsection{Double Q-learning for Trajectory Optimization}
Given the components of the double Q-learning algorithm, we explain how to use the double Q-learning algorithm to solve the problem in \eqref{eq:max}. To find the optimal trajectory, the proposed learning algorithm needs to use Q-tables to store the reward values resulting from different states and actions.  
 
In contrast to the traditional Q-learning that determines the action selection policy and the values of Q-table using one Q-table, the proposed algorithm uses two Q-tables to separately determine the action selection policy and update the values of Q-table. Hence, the proposed algorithm can avoid the overestimation in Q-learning. The optimal Q values can be given by \textit{Bellman's optimal equation} \cite{NIPS2010_3964}:
 \begin{equation}\begin{aligned}
  \color{black}
  Q^*\left(\boldsymbol s,a\right)=\sum_{\boldsymbol s'}p_{\boldsymbol s\boldsymbol s'}\left(r\left(\boldsymbol s,a\right)+\gamma\max_{a'}Q^*\left(\boldsymbol s',a'\right)\right),\label{eq:Q}
 \end{aligned}\end{equation}
where $\gamma$ is the discount factor, and $\color{black}{p_{{\boldsymbol{s}}{\boldsymbol{s}'}}}$ is the probability from state $\color{black}\boldsymbol{s}$ to $\color{black}\boldsymbol{s}'$. To enable {\color{black} the UAV} to record the  values of the Q-tables in \eqref{eq:Q}, the Q-tables need to update at each time slot $t$, which can be given as follows:
 \begin{equation}\color{black}
\begin{aligned}
  Q_{t+1}^A\left(\boldsymbol s,a\right)=& \left(1-\alpha_\textrm{L}\right)Q_{t}^A\left(\boldsymbol s,a\right)\\
  &+\alpha_\textrm{L}\left(r\left(\boldsymbol s,a\right)  +\gamma Q^B\left(\boldsymbol s',a^*\right)\right),\label{eq:qa}
\end{aligned}\end{equation}
\begin{equation}\color{black}
\begin{aligned}
  Q_{t+1}^B\left(\boldsymbol s,a\right)=& \left(1-\alpha_\textrm{L}\right)Q_{t}^B\left(\boldsymbol s,a\right)\\
  &+\alpha_\textrm{L}\left(r\left(\boldsymbol s,a\right)  +\gamma Q^A\left(\boldsymbol s',b^*\right)\right),\label{eq:qb}
\end{aligned}\end{equation}
where $\alpha_\textrm{L}$ is the learning rate, $\color{black}a^*=\arg\max_{a'}Q^A\left(\boldsymbol s',a'\right)$, $\color{black}b^*=\arg\max_{a'}Q^B\left(\boldsymbol s',a'\right)$, and $\color{black}\boldsymbol s'$ is the next state after taking action $\color{black}a$ at state $\color{black}s$. Note that, at each iteration, only one Q-table will be updated. To update the values of Q-tables in \eqref{eq:qa} or \eqref{eq:qb}, {\color{black} the UAV} needs to select one action to implement at time $t$. The action selection policy of {\color{black} the UAV} is given by:  
\begin{equation}\label{eq:p}
\color{black}
\pi _{{a}} \!=\! \left\{ {\begin{array}{*{20}{c}}
{1 - \varepsilon  + \frac{\varepsilon }{{\left| {{\mathcal{A}}} \right|}},\;\; \arg\! \mathop {\max }\limits_{{{a}} \in {\mathcal{A}}}\! Q\left(\boldsymbol{s},a\right),}\\
{\;\;\;\;\;\;\;\frac{\varepsilon }{{\left| {{\mathcal{A}}} \right|}},\;\;\;\;\;\;\;\;\text{otherwise},\;\,\,\,\;\;\;\;\;\;\;\;\;}
\end{array}} \right. 
\end{equation}
where $\pi _{a}$ denotes the probability that {\color{black} the UAV} takes action $a$, $\varepsilon$ is the exploration probability of a random action, ${\mathcal{A}}$ denotes the set of actions of {\color{black} the UAV}, and $\left| {{\mathcal{A}}} \right|$ is the total number of actions of {\color{black} the UAV}. Note that, the proposed algorithm has two Q-tables. Hence, one of the Q-table is used to determine the action selection policy in \eqref{eq:p} and the other Q-table is used to update the value of Q-table in \eqref{eq:qa} or \eqref{eq:qb}. For example, if $Q_{t+1}^A\left(\boldsymbol s,a\right)$ is used to determine the action selection policy, then $Q_{t+1}^B\left(\boldsymbol s,a\right)$ must be used to determine the value of Q-table. Based on the above formulations, the double Q-learning algorithm performed by {\color{black} the UAV} is summarized in Algorithm \ref{al:tra}.

\subsection{Convergence of the Proposed Algorithm}
Next, we analyze the convergence of the proposed double Q-learning algorithm. We first prove that the proposed framework is an Markov Decision Process (MDP) \cite{Abu2015MDP}. Then, we prove that the proposed algorithm will converge and find the optimal trajectory for the UAV to maximize the number of satisfied users.


The following theorem proves that the proposed framework is an MDP, which is given by:
\newtheorem{theorem}{\bf Theorem}
\begin{theorem}\label{theorem:MDP}
The proposed double Q-learning framework is an MDP.
\end{theorem}

\begin{algorithm}[h]
\caption{Double Q-learning based algorithm for trajectory optimization}
\label{al:tra}
\begin{small}
\begin{algorithmic}[1]
\STATE \textbf{Input:} User positions and user requests
\STATE \textbf{Init:} UAV position, $Q^A$, $Q^B$, $\color{black}\boldsymbol s$
\REPEAT
\IF{$rand\left(\cdot\right)<\varepsilon$}
\STATE [exploration step]
\STATE randomly choose one action
\STATE receive immediate reward for the agent UAV
\STATE Update table $Q^A$ or $Q^B$ as given in \eqref{eq:qa} \eqref{eq:qb}
\ELSE
\STATE [exploitation step]
\STATE Choose (e.g. in turn, randomly) either update table $Q^A$ or table $Q^B$
\IF{update table $Q^A$}
\STATE Choose action $\color{black}a=\arg \max_{a}Q^B\left(\boldsymbol s,a\right)$ from $Q^B$\STATE receive immediate reward for the agent UAV
\STATE Update table $Q^A$ as given in \eqref{eq:qa}
\ELSIF{update table $Q^B$}
\STATE Choose action $\color{black}a=\arg \max_{a}Q^A\left(\boldsymbol s,a\right)$ from $Q^A$\STATE receive immediate reward for the agent UAV
\STATE Update table $Q^B$ as give in \eqref{eq:qb}
\ENDIF
\ENDIF
\STATE $\color{black}\boldsymbol s\leftarrow\boldsymbol s'$
\UNTIL{end}
\end{algorithmic}
\end{small}
\end{algorithm}
\begin{IEEEproof}
  An MDP consists of five basic components \cite{Abu2015MDP}: 1) a finite set of states, 2) a finite set of actions, 3) a transition probability function, 4) the immediate reward function, and 5) the set of decision epoch, which can be finite or infinite. Next, we prove that the components of the proposed double Q-learning framework satisfied the conditions of an MDP. 

  In the proposed framework: 1) The number of actions is equal to the number of users. In consequence, the set of actions is limited. 2) The number of states is equal to the $U$-th power of 2, which indicates all the possible combinations of whether each user has been served. Thus, the set of states is limited. 3) The action selection policy \eqref{eq:p} indicates the probability of {\color{black} the UAV} taking action $\color{black}a$, which also determines the transition probability from current state $s$ to next state $s'$. 4) The reward function is immediately determined by the current state and the action to be taken. 5) {\color{black} the UAV} takes decisions at any time, which leads to an infinite decision epoch.

  In summary, the proposed double Q-learning framework satisfied all the conditions of an MDP. Therefore, the framework is an MDP.
\end{IEEEproof}

From Theorem~\ref{theorem:MDP}, we can see that the proposed double Q-learning framework is an MDP with five basic components. Thus, the convergence of the proposed Q-learning algorithm can be viewed as the convergence of an MDP, which is given by the following corollary.

\newtheorem{corollary}{\bf Corollary}
\begin{corollary}\label{cor:converge}
  In the proposed double Q-learning algoritm, both $Q^A$ and $Q^B$ will converge to the optimal value function $Q^*$ with probability one eventually. 
\end{corollary}

\begin{IEEEproof}
  The work in \cite[Theorem 1]{NIPS2010_3964} proved that, for an MDP corresponding to a double Q-learning algorithm, both Q-tables converge to the same optimal value under the following conditions: 1) the MDP is finite, i.e. $\left|\mathcal S\times\mathcal A\right|<\infty$; 2) the discount factor $\gamma \in \left[0,1\right)$; 3) the Q values are stored in a lookup table; 4) both $Q^A$ and $Q^B$ receive an infinite number of updates; 5) the learning rate $\alpha_\textrm{L} \in \left[0,1\right]$; 6) the reward function is finite. 

  In the proposed framework: 1) Both states and actions are finite. In consequence, the MDP is finite. 2) $\gamma$ is set to a reasonable value in $\left[0,1\right)$. 3) Two Q-tables store all the Q values related to states and actions. The Q values can be looked up by the state and action. 4) Both $Q^A$ and $Q^B$ can be updated infinitely without artificial limits. 5) $\alpha_\textrm{L}$ is set to a reasonable value in $\left[0,1\right]$. 6) The reward function represents the number of satisfied users taking action $a_j$ at current state $\boldsymbol s_j$. Thus, the result is an integer less than the number of total users $U$. Obviously, the reward function is finite. 

  In consequence, the proposed algorithm satisfies all of the conditions in \cite[Theorem 1]{NIPS2010_3964}. Thus, both $Q^A$ and $Q^B$ will converge to the same optimal value.
\end{IEEEproof}

From Corollary~\ref{cor:converge}, we can see that both $Q^A$ and $Q^B$ in the proposed algorithm can converge to the same optimal value. The optimal value corresponds to the optimal trajectory which leads to the maximum of satisfied users. Therefore, as the proposed algorithm converges, it can maximize the number of satisfied users.

\section{Simulation Results}\label{sec:sim}
In our simulations, we consider a circular UAV based cellular network area with a radius $r = 200$ m, $U=20$ uniformly distributed users and a UAV. The number of ground users is equal to the number of aerial users, $U_{\rm G}=U_{\rm A}$. For implementing the proposed double Q-learning algorithm, we use the Matlab tools. Other system parameters are listed in Table \ref{table:sim_para}. We compare the proposed algorithm with a random algorithm that selects the user to serve in a random order and the traditional Q-learning algorithm in \cite{Bennis2010Q}. All statistical results are averaged over 5000 independent runs. 
\begin{table}[h]
\caption{System parameters}
\label{table:sim_para}
\centering
\begin{small}
\begin{tabular}{c|c|c}
\hline
\textbf{Parameters} & \textbf{Description} & \textbf{Values}\\
\hline\hline
$H$ & UAV altitude & 100 m\\
\hline
$\alpha$ & Path loss exponent & 2\\
\hline
$\eta_{\rm NLoS}$ & NLoS attenuation factor & 0.3\\
\hline
$\eta_{\rm LoS}$ & LoS attenuation factor & 2\\
\hline
$X,Y$ & Environment parameters & 11.95, 0.136\\
\hline
$\sigma^2$ & Noise power &-74 dBm\\
\hline
$P$ & UAV transmit power & 5 W\\
\hline 
$B$ & Bandwidth & 1 MHz\\
\hline
$f$ & mmWave frequency & 35 GHz\\
\hline
$c_{\rm U}$ & UAV speed & 50 m/s\\
\hline
$\alpha_\textrm{L}$ & Learning rate & 0.5\\
\hline
$\gamma$ & Discount rate & 0.8\\
\hline
$\varepsilon$ & Exploration rate & 0.5\\
\hline
\end{tabular}
\end{small}
\end{table}

\begin{figure}[!t]
\centerline{\includegraphics[width=3in]{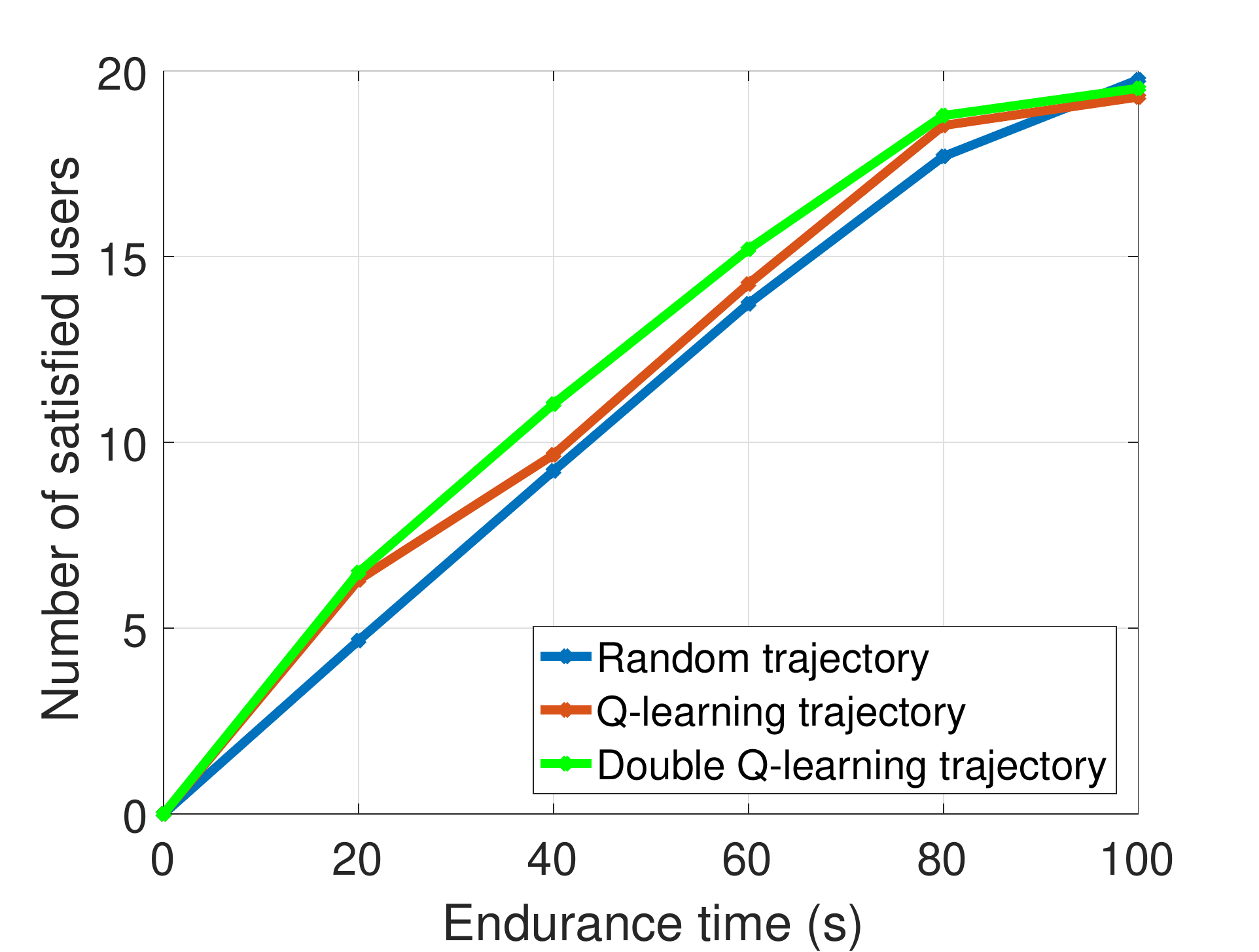}}
\caption{Number of satisfied users as endurance time varies.}
\label{figure:time}
\end{figure}

In \figurename~\ref{figure:time}, we show how the number of satisfied users changes as the endurance time changes. From \figurename~\ref{figure:time}, we can see that, as the endurance time increases, the number of satisfied users increases. This is due to the fact that the UAV can complete more user requests at longer user's endurance time. \figurename~\ref{figure:time} also shows that the number of satisfied users in all three algorithms achieves up to 20 when the endurance time increases to 100. This is due to the fact that, as the endurance time is long enough, the UAV can complete all the user requests in arbitrary trajectory.

\begin{figure}[!t]
\centerline{\includegraphics[width=3in]{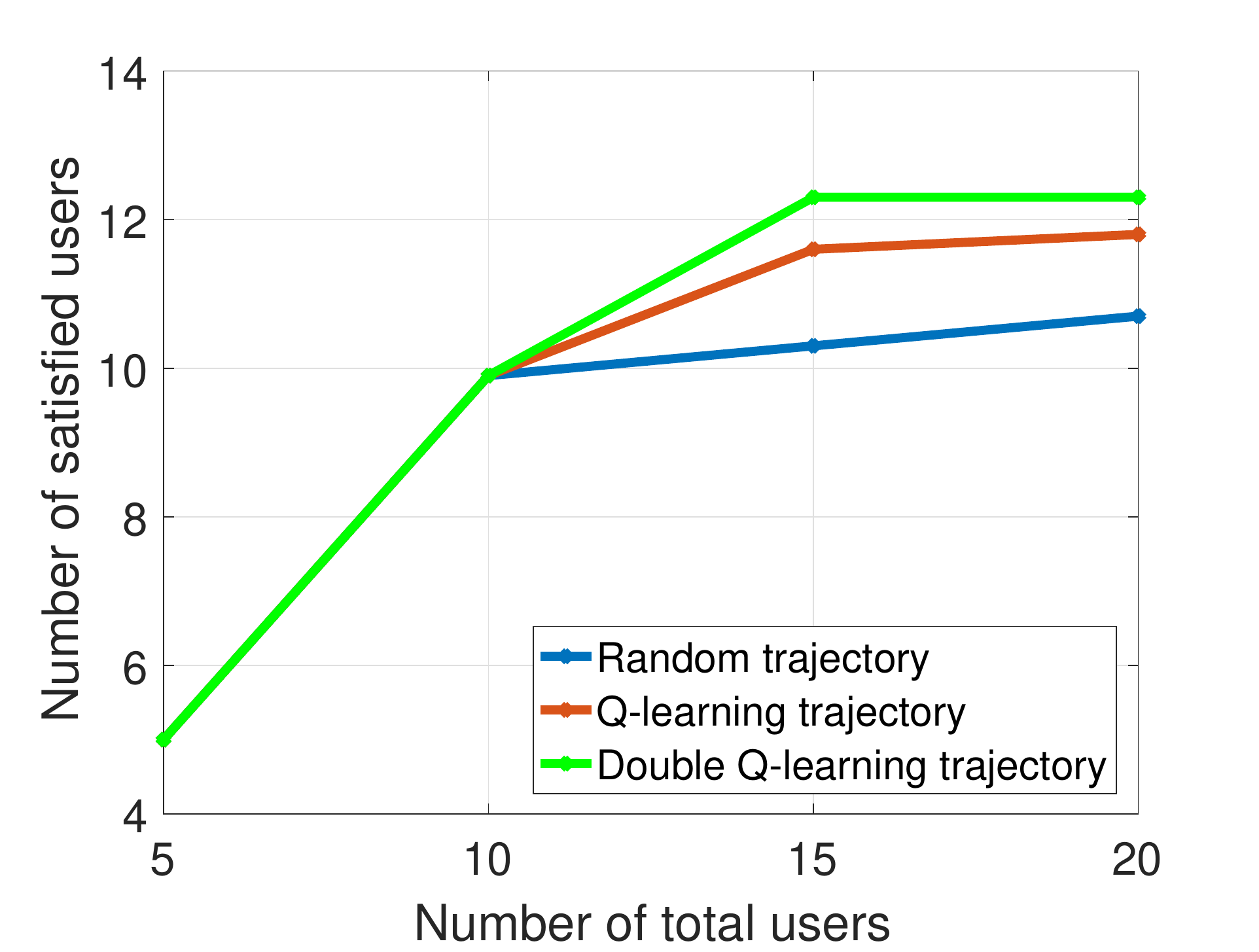}}
\caption{Number of satisfied users as the number of total users varies.}
\label{figure:user}
\end{figure}

\figurename~\ref{figure:user} shows how the number of satisfied users changes as the total number of users varies, as endurance time is 50 seconds. {\color{black}The simulation results are averaged over 5000 independent runs, and, hence, the number of satisfied users in \figurename~\ref{figure:user} is not an integer.} From \figurename~\ref{figure:user}, we can see that, when the number of users is small such as $U=5$, the number of satisfied users is equal to the number of total users. This is due to the fact that the user's endurance time is long enough so that the request of any user can be completed within the endurance time in arbitrary trajectory. In \figurename~\ref{figure:user}, we can also see that the number of satisfied users increases till $U=15$. This is due to the fact that the UAV can serve more users within the endurance time. Meanwhile, \figurename~\ref{figure:user} also shows that the number of satisfied users remains almost unchanged when the number of total users increases from 15 to 20. This is due to the fact that the UAV can only serve a limited number of users due to the endurance time requested by each user. \figurename~\ref{figure:user} also indicates that the proposed algorithm can achieve up to 19.4\% and 14.1\% gains in terms of the number of satisfied users compared to the random algorithm and Q-learning algorithm, respectively. These gains stem from the fact that the proposed algorithm aims to find the optimal trajectory to maximize the number of satisfied users, while the random algorithm only gives a trail random trajectory regardless of the number of satisfied users, and the Q-learning algorithm may result in sub-optimal policies which lead to a worse result.

\begin{figure}[!t]
\centerline{\includegraphics[width=3in]{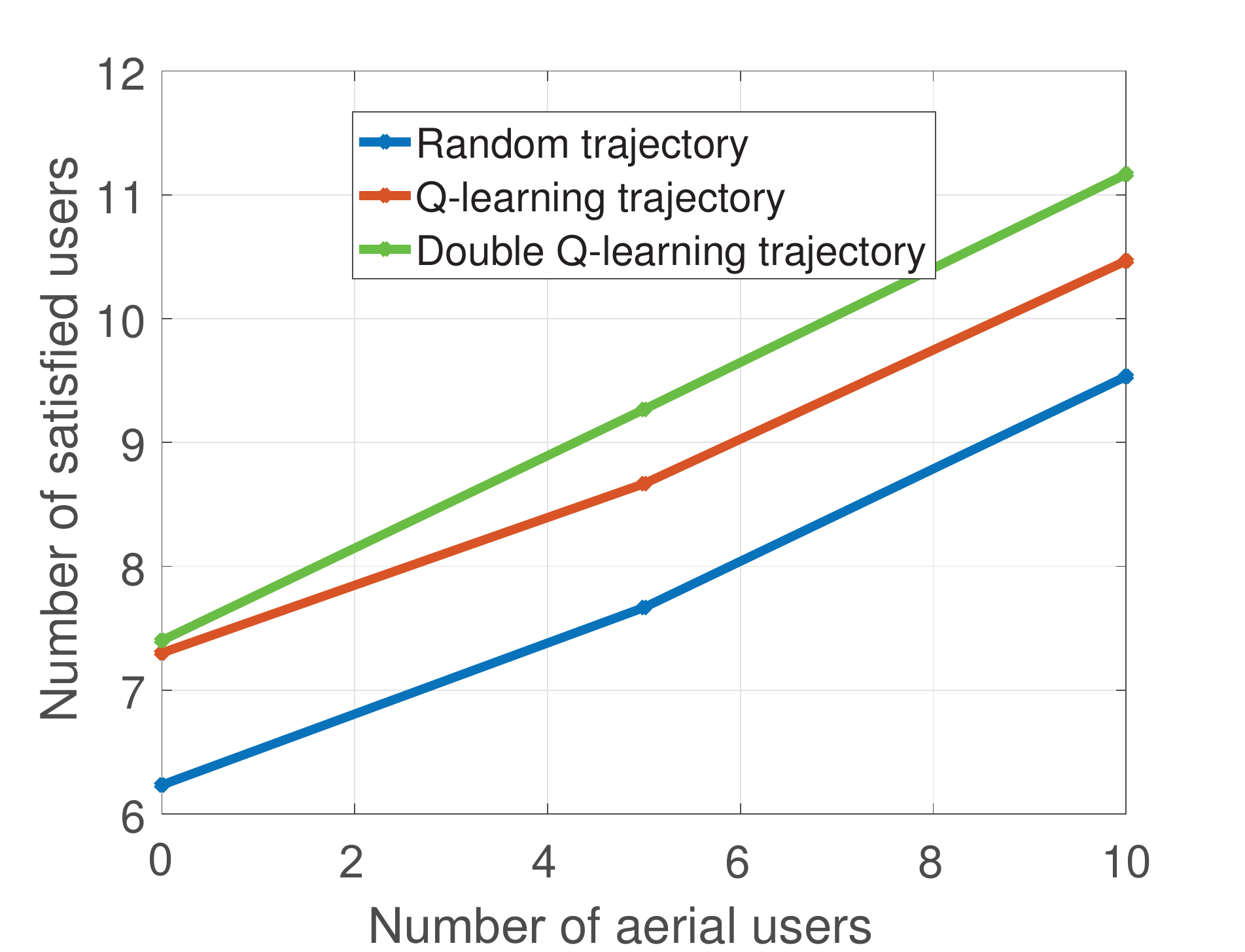}}
\caption{Number of satisfied users as the number of aerial users varies.}
\label{figure:usertype}
\end{figure}

In \figurename~\ref{figure:usertype}, we show how the number of satisfied users changes as the number of aerial users changes.
From \figurename~\ref{figure:usertype}, we can see that, as the number of aerial users increases, the number of ground and aerial users that are satisfied with the service provided by the UAV increases. This is due to the fact that the channel condition of the aerial users is better than that of the ground users. In consequence, the UAV can spend less time to serve an aerial user.
\figurename~\ref{figure:usertype} also shows that, in terms of the number of satisfied users, the proposed algorithm can achieve up to 20.1\% and 6.7\% gains compared to the random algorithm and Q-learning algorithm, respectively.

\begin{figure}[!t]
\centerline{\includegraphics[width=3in]{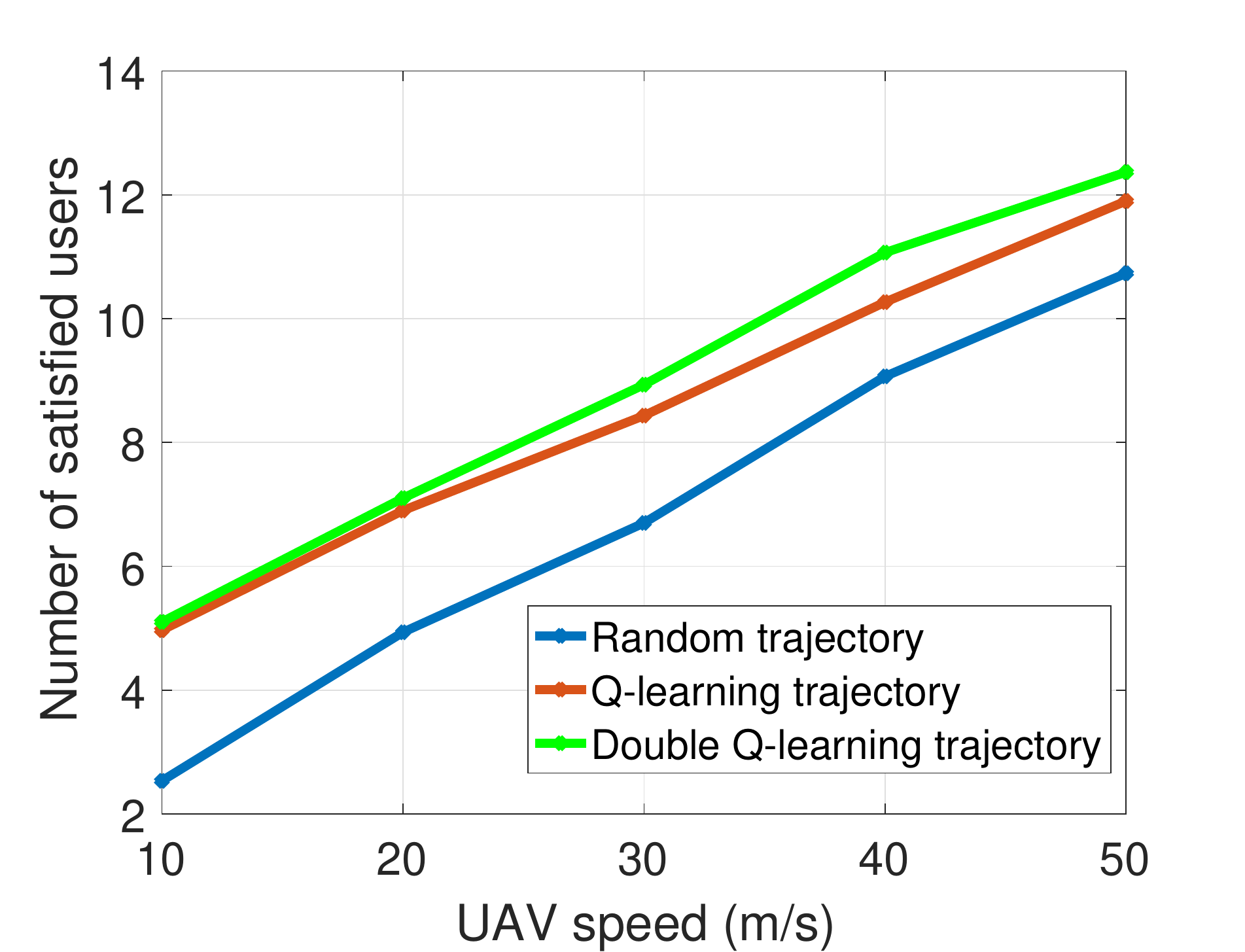}}
\caption{Number of satisfied users as UAV speed varies.}
\label{figure:speed}
\end{figure}

\figurename~\ref{figure:speed} shows how the number of satisfied users changes as the UAV speed changes. From \figurename~\ref{figure:speed}, we can see that, as the UAV speed increases, more users are satisfied with the service provided by the UAV. This is due to the fact that the faster speed leads to the shorter flying time. In consequence, the UAV spends more time to serve users rather than to fly in the air.
\figurename~\ref{figure:speed} also shows that the proposed algorithm can achieve up to 22.1\% and 7.8\% gains in terms of the number of satisfied users compared to the random algorithm and Q-learning algorithm. These gains stem from the fact that the flying time, which is inversely proportional to the UAV speed, accounts for a great part of the total service time. The proposed algorithm can find the optimal trajectory which saves the flying time of the UAV, while the random trajectory doesn't concern about the flying time and the Q-learning algorithm may work out a sub-optimal trajectory which wastes the service time on flying.

\begin{figure}[!t]
\centerline{\includegraphics[width=3in]{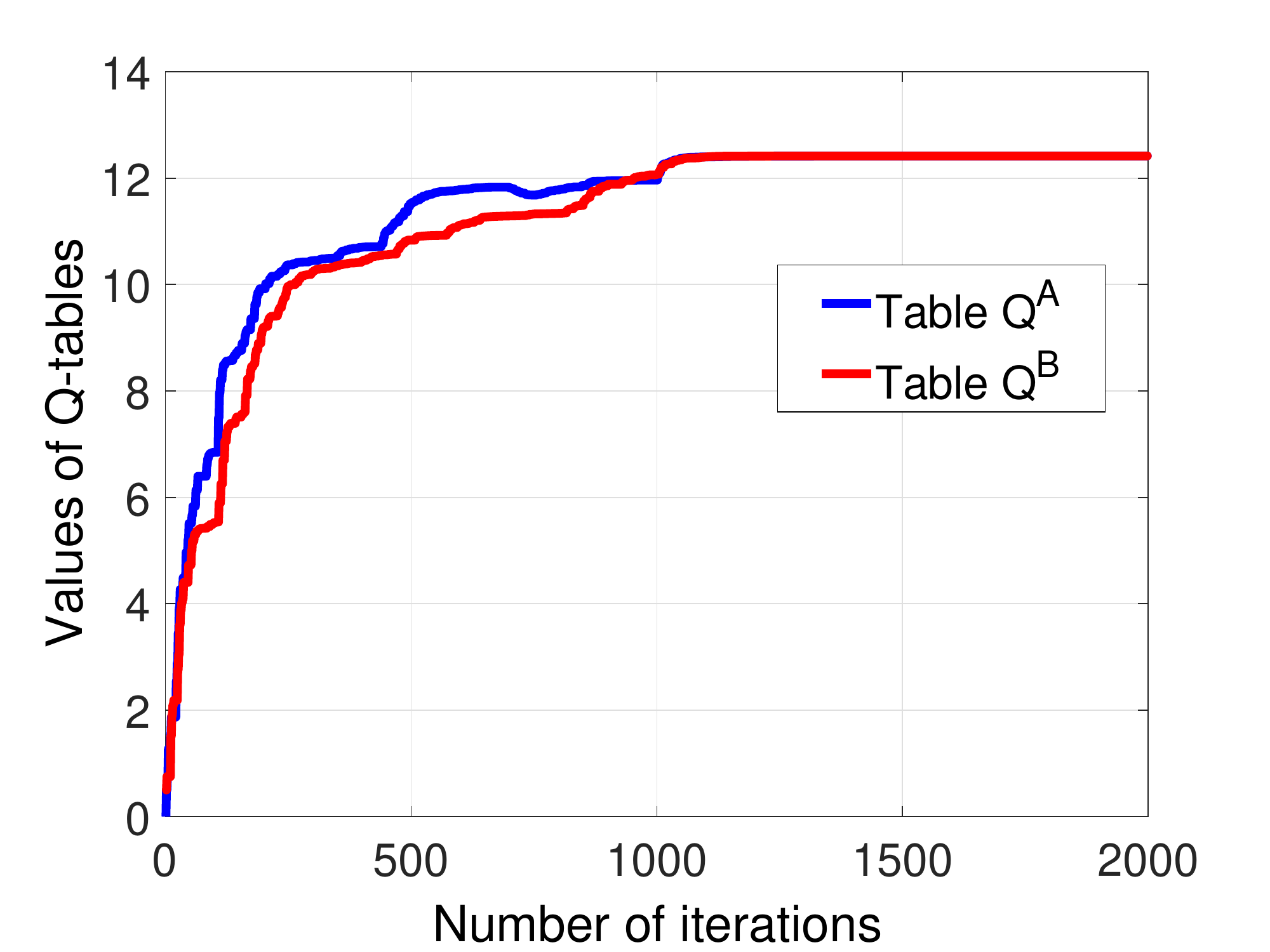}}
\caption{Convergence of the proposed double Q-learning algorithm.}
\label{figure:convergence}
\end{figure}

\figurename~\ref{figure:convergence} shows the number of iterations needed till convergence for the proposed double Q-learning approach. In this figure, we can see that, as time elapses, the values of Q-tables increase until convergence to their final values. \figurename~\ref{figure:convergence} also shows that the proposed approach needs 1000 iterations to reach convergence. From \figurename~\ref{figure:convergence}, we can also see that, tables $Q^A$ and $Q^B$ may have different values as time elapses. However, as time continues to elapse, tables $Q^A$ and $Q^B$ will converge to the same final value. This is due to the fact that, at each iteration, the proposed double Q-learning algorithm selects an action based on the value of one Q-table and updates the action's Q-value in another Q-table. The result also confirms Corollary~\ref{cor:converge}.

\begin{figure}[!t]
\centerline{\includegraphics[width=3in]{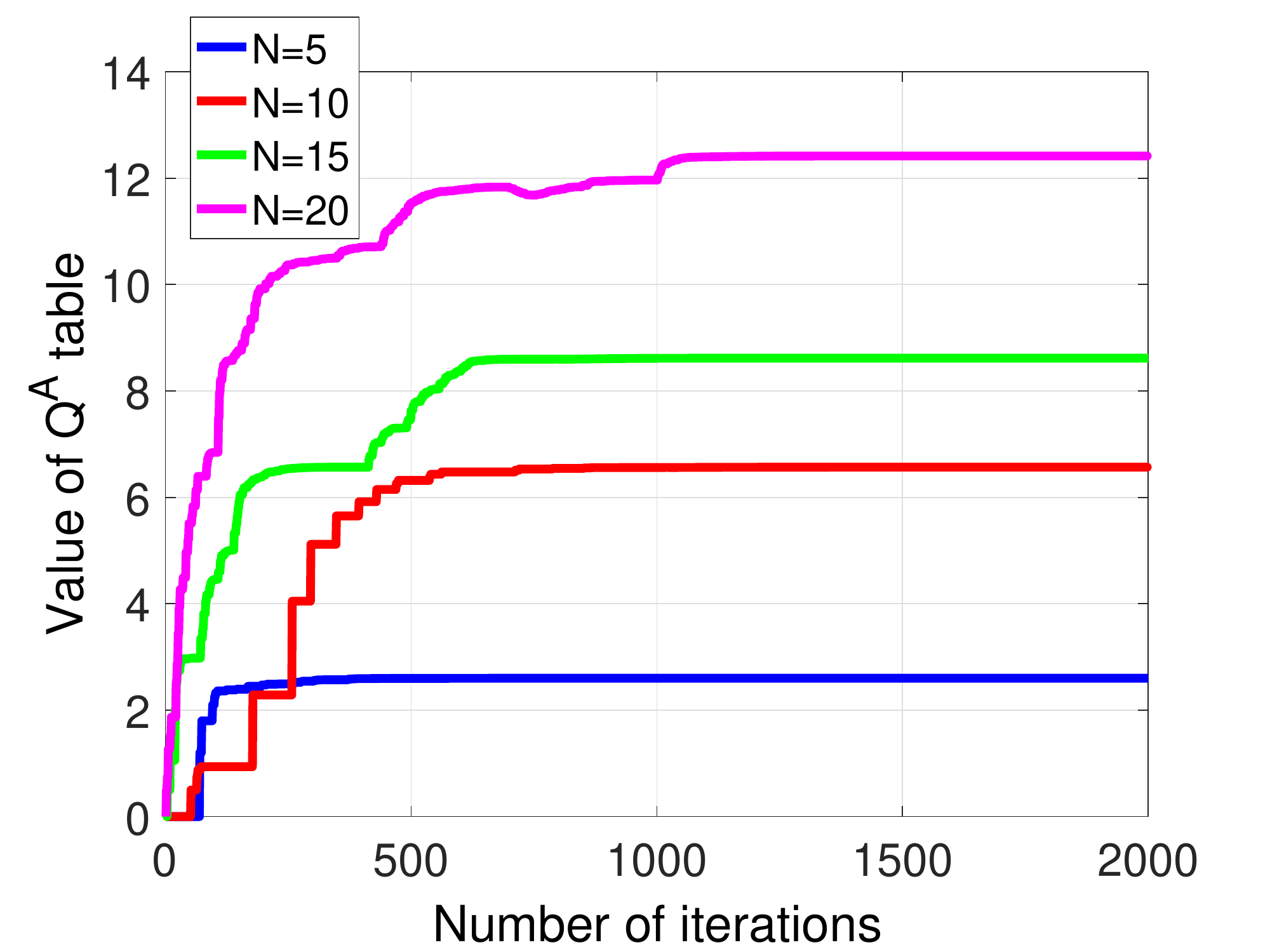}}
\caption{Convergence as the number of total users varies.}
\label{figure:con_compare}
\end{figure}

\figurename~\ref{figure:con_compare} shows how the convergence changes as the number of total users varies. From \figurename~\ref{figure:con_compare}, we can see that, as the number of users increases from 5 to 20, the proposed approach needs 100, 500, 600, 1000 iterations to reach convergence, respectively. This is due to the fact that, as the number of total users increases, the number of states of the proposed double Q-learning framework increases exponentially. In consequence, the proposed algorithm needs more iterations to explore those states and, hence, it uses more iterations to reach the convergence and find the optimal trajectory.

\begin{figure}[!t]
\centerline{\includegraphics[width=3in]{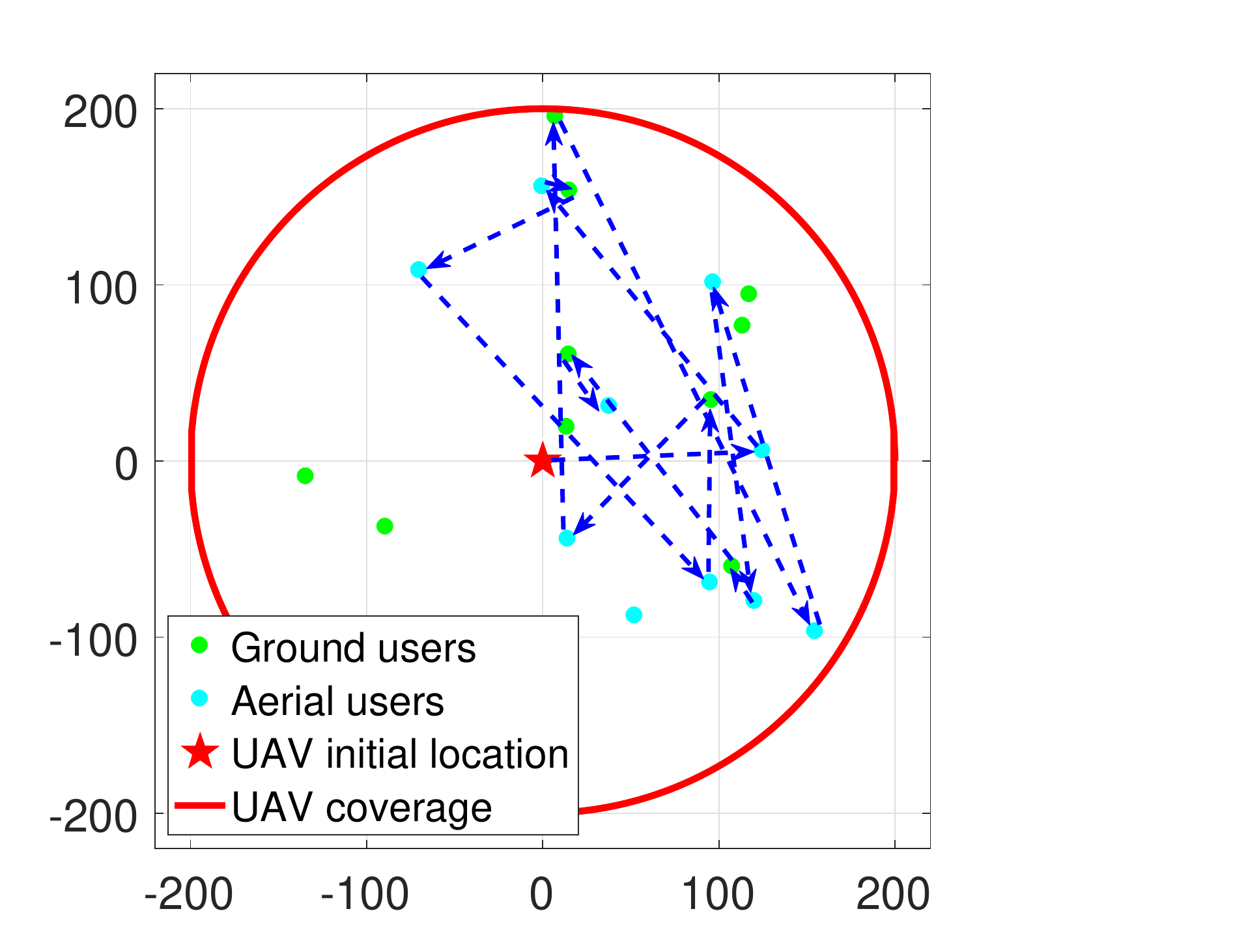}}
\caption{An illustrative example of the UAV trajectory maximized by the proposed algorithm.}
\label{figure:trajectory}
\end{figure}

\figurename~\ref{figure:trajectory} shows an optimal trajectory example designed by the double Q-learning framework for a network with ground users and aerial users as the blue arrow being the UAV trajectory. In this figure, the UAV starts from the origin of the coordinates and then selects the users to serve. From \figurename~\ref{figure:trajectory}, we can also see that some of the users are not served by the UAV, this is because their requests cannot be completely served before their endurance time. \figurename~\ref{figure:trajectory} also shows that more aerial users than ground users are served by the UAV. This is due to the fact that the aerial users have better channel conditions and faster transmission rates. In consequence, the UAV is more willing to serve an aerial user than a ground user.

\section{Conclusion}\label{sec:con}
In this paper, we have developed a novel framework that enables flying, movable UAVs to provide service for the three dimensional users in a cellular network. We have formulated an optimization problem that seeks to maximize the number of satisfied users. To solve this problem, we have developed a novel algorithm based on the machine learning tools of double Q-learning. The proposed algorithm enables the UAVs to find the optimal flying trajectory so as to maximize the number of satisfied users. Simulation results have shown that the proposed approach yields significant performance gains in terms of the number of satisfied users compare to random algorithm and Q-learning algorithm.
\vskip1.55cm

\bibliographystyle{IEEEtran}


\end{document}